# Noise optimization of the source follower of a CMOS pixel using BSIM3 noise model

Swaraj Mahato[a], Guy Meynants[b], Gert Raskin[a], J. De Ridder[a], H. Van Winckel[a]
[a]Institute of Astronomy, KU Leuven, Celestijnenlaan 200D, 3001 Leuven, Belgium; [b] CMOSIS nv, Coveliersstraat 15, 2600 Antwerp, Belgium

## ABSTRACT

CMOS imagers are becoming increasingly popular in astronomy. A very low noise level is required to observe extremely faint targets and to get high-precision flux measurements. Although CMOS technology offers many advantages over CCDs, a major bottleneck is still the read noise. To move from an industrial CMOS sensor to one suitable for scientific applications, an improved design that optimizes the noise level is essential. Here, we study the 1/f and thermal noise performance of the source follower (SF) of a CMOS pixel in detail. We identify the relevant design parameters, and analytically study their impact on the noise level using the BSIM3v3 noise model with an enhanced model of gate capacitance. Our detailed analysis shows that the dependence of the 1/f noise on the geometrical size of the source follower is not limited to minimum channel length, compared to the classical approach to achieve the minimum 1/f noise. We derive the optimal gate dimensions (the width and the length) of the source follower that minimize the 1/f noise, and validate our results using numerical simulations. By considering the thermal noise or white noise along with 1/f noise, the total input noise of the source follower depends on the capacitor ratio $C_G/C_{FD}$ and the drain current ($I_d$). Here, $C_G$ is the total gate capacitance of the source follower and $C_{FD}$ is the total floating diffusion capacitor at the input of the source follower. We demonstrate that the optimum gate capacitance ($C_G$) depends on the chosen bias current but ranges from $C_{FD}/3$ to $C_{FD}$ to achieve the minimum total noise of the source follower. Numerical calculation and circuit simulation with 180nm CMOS technology are performed to validate our results.

## 1. INTRODUCTION

Traditionally, in scientific applications such as astronomy and astrophysics CCDs have been used as the detector of choice for their excellent performance with high quantum efficiency (QE) and low read noise. Due to rapid and significant technology advancement over the last decade CMOS detectors are becoming increasingly popular in astronomy [1], [2]. This is because CMOS detectors offer number of inherent advantages compare to CCDs in terms of low cost, low power, flexible readout capability and the potential to do integrated signal processing. Although CMOS sensors are rapidly overtaking the commercial market, a major bottleneck is still the read noise [3] for scientific imaging. The night sky offers an abundant number of field-of-views with a huge range in brightness (e.g. dense stellar clusters). It also shows coherent brightness variations. In astronomy, observing very close pairs of stars with a large brightness ratio demands high dynamic range of the detector. One would like to reduce the noise level to increase the dynamic range of the image sensor.

One of the most common objectives in CMOS sensor design for scientific application is to minimize the equivalent noise charge (ENC). During last few years, different circuit techniques effectively reduce the contribution of pixel offset, reset noise and thermal noise. However, 1/f noise, originated from in-pixel source follower still dominates the read noise [4], [5]. Usually optimum gate dimension of the source follower is chosen as the possible design solution to reduce 1/f noise. In existing literatures most of the works on the noise optimization and noise analysis have been done with over simplified classical noise models but they are not in agreement with simulator noise models [6]. Identifying the exact association between the geometry of the source follower, technology parameter (i.e. the oxide thickness) and the input referred noise charge, gives some room to do better optimization. We understand that the current spice simulator noise models need to be analytically examined to achieve accuracy and agreement. In this work, the 1/f and thermal noise performance of the source follower of a CMOS pixel are studied in detail with advanced simulation models.

*Email: swarajbandhu.mahato@ster.kuleuven.be

In this paper Section 2 presents the analog readout chain of the typical CMOS image sensor and discusses the effect of correlated double sampling (CDS) on 1/f noise. Analytical 1/f noise calculation and optimization is presented in Section 3. There, the analysis has been started with classical noise model. Then, gradually we have extended our study towards the enhanced analysis of 1/f noise of the source follower (SF) for deep submicron technology with BSIM3v3 simulation model. In Section 4, analysis of total noise (1/f noise and thermal noise) of the source follower (SF) and the optimization technique have been discussed. We have simulated CMOS pixel circuit with current available technology and results have been presented in Section 5. This is followed by the conclusion of our discussion in Section 6.

## 2. READOUT CHAIN AND EFFECT OF CDS ON 1/F NOISE

The typical readout chain contains a 4T CMOS pixel with pinned photodiode (PPD), column level pre-amplifier and correlated double sampling (CDS) or sample-and-hold circuits. Figure 1 shows the schematic diagram of the readout chain.

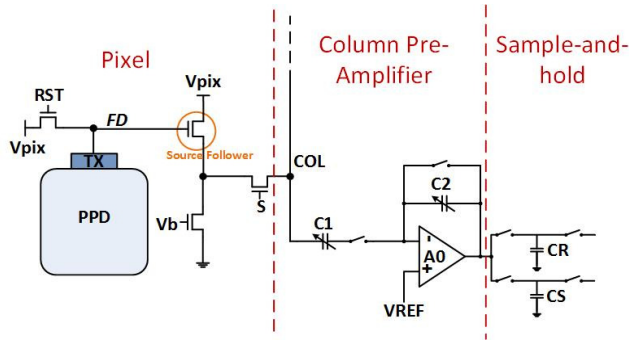

Figure 1: Readout chain of a CMOS image sensor with 4T CMOS pixel, column pre-amplifier and CDS

When row selector switch (S) is 'ON' the sense node (FD) is reset to a voltage $V_R$. After auto-zeroing of the column pre-amplifier, the reset voltage level is sampled. Then, the transfer gate (TX) is 'ON' to allow the generated charge from PPD to sense node. Signal voltage ($V_S$) at sense node is then sampled and finally both samples (reset and signal) are differentiated. It is well understood that 1/f noise of the source follower has significant contribution to the read noise of the CMOS image sensor (CIS). Even with applying CDS technique, 1/f noise, generated from in-pixel source follower dominates the read noise. To analyze the 1/f noise in CIS we first discuss the CDS effect on 1/f noise.

CDS as a noise reduction technique is indispensable in image sensor to reduce kTC noise, as well as fixed pattern noise (FPN) and 1/f noise [7], [8]. Sample-and-hold circuits are typically used to implement the CDS operation as depicted in Figure 1. More generic form of CDS is called as correlated multiple sampling (CMS). In CMS both the reset and detected signal levels are sampled for multiple times. Then, the difference of the average of the both levels is calculated to do pixel-related noise cancelation. The sampling timing of CMS operation is shown in Figure 2, where $T_0$ is the sampling period.

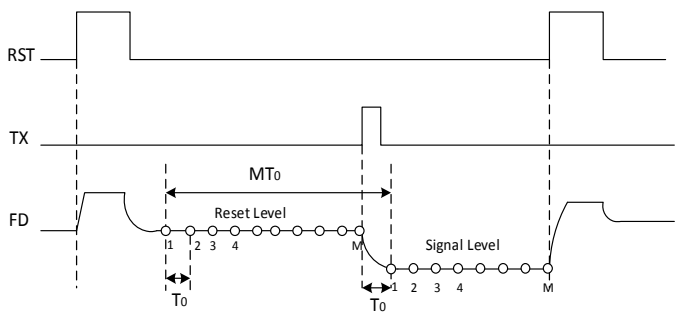

Figure 2: Timing diagram of CMS operation.

In other words, CMS technique results the average of the voltage signals after successive CDS operations, which is expressed by,

$$y = \frac{1}{M}\sum_{i=1}^{M}\{V_R(i) - V_S(i)\} \quad (1)$$

Where, $M$ is the total number of sampling, $V_R(i)$ and $V_S(i)$ are the detected reset level and signal level respectively for the $i^{th}$ sample. In the readout chain signal is low-pass filtered before sampling. By considering a 1st order low pass filter with the cutoff angular frequency $\omega_c$, 1/f noise power of CMS after M times sampling is given by[9],

$$P(M) = \frac{1}{M^2}\int_0^\infty S_n(x)_{\frac{1}{f}} \cdot \frac{1}{1+(x/x_c)^2}\frac{4\sin^4 x}{\sin^2(x/M)} dx \quad (2)$$

Here, $x_c = \omega_c M T_0/2$ and $S_n(x)_{1/f} = \frac{N_f}{f}$ is the power spectral density (PSD) of the input 1/f noise source.

Now for CDS operation we take one sample for each signal so, M=1. Then the 1/f noise power for CDS is given by,

$$P(1) = \int_0^\infty S_n(x)_{1/f} \frac{4\sin^2 x}{1+(x/x_c)^2} dx \quad (3)$$

Eqs.3 corresponds to the "double delta" [10] and the special case of "differential average" [10]. Then the solution for 1/f noise with M=1 is,

$$P(1) \cong 2N_f\{C + \ln(\omega_c T_0)\} \quad (4)$$
$$P(1) \cong \beta_{CDS} N_f \quad (5)$$

Where $C = 0.577215\ldots$ is Euler's Constant and $\beta_{CDS} \cong 2\{C + \ln(\omega_c T_0)\}$. From Eqs.4 we can see that 1/f noise power after CDS operation is a function of $\omega_c T_0$. Based on the numerical calculation of Eqs.4, Figure 3 shows the normalized 1/f noise power as a function of $\omega_c T_0$.

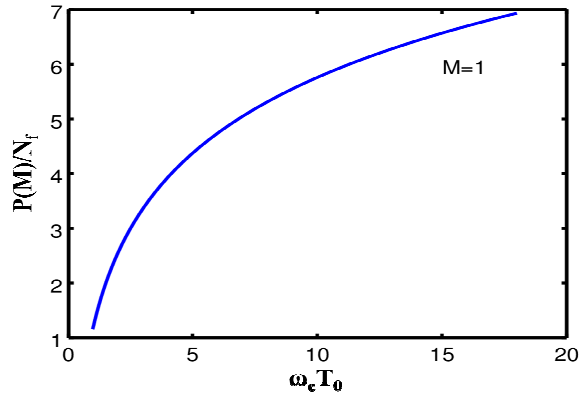

Figure 3: Calculated 1/f noise power after CDS (M=1) as a function of $\omega_c T_0$.

In usual design steps of readout circuit for CMOS image sensor, the value of $\omega_c T_0$ is chosen such that to achieve precise cancellation of the fixed pattern noise (FPN). For example, with $\omega_c T_0 \cong 7$, the settling error is significantly low (~0.1%) and at that point from Figure 3 we can get $P(1)/N_f \cong 5$. This arbitrary value is used for further numerical calculation as in examples.

## 3. 1/F NOISE ANALYSIS OF THE IN-PIXEL SOURCE FOLLOWER

In the following analysis noise source of the source follower is assumed as statistically independent and uncorrelated to the other noise sources. Our analysis starts with classical noise model and extended to the advanced BSIM model, what is indeed the main focus of this work.

### 3.1 1/f noise analysis with classical models and optimization

The in-pixel source follower (MOSFET) needs to be optimized to get minimum mean-square noise charge contribution. Input referred equivalent 1/f noise charge of the MOSFET depends on total physical capacitance associated to its input.

Figure 4 shows the pixel structure with capacitance, connected to the input of the source follower (M2). Typically this total input capacitance is set by the sensor elements and interconnects.

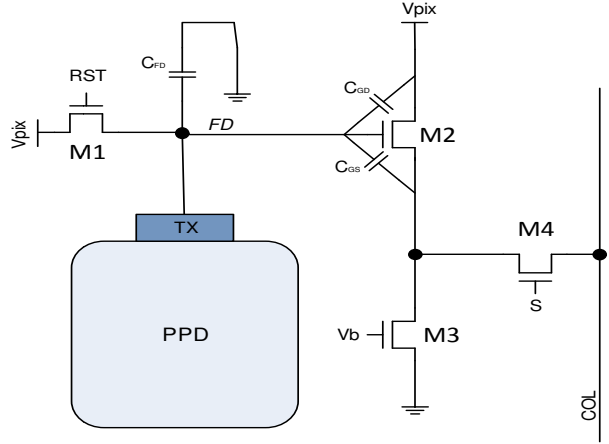

Figure 4: CMOS pixel with capacitance at FD node.

Total physical capacitance connect to the gate of M2 (source follower), shown in Figure 4 is,

$$C_{total} = C_{FD} + C_{GS} + C_{GD} = C_{FD} + C_{gs}WL + C_{gd}W \qquad (6)$$

Here, $C_{FD}$ is the floating diffusion capacitor of the pixel at sense node. This $C_{FD}$ includes junction capacitance of the reset transistor (M1) drain, the gate source overlap capacitances of the transfer gate (TX) and M1, and the parasitic capacitance of the interconnect routing. *Cgs* is the gate-source capacitance per unit area, *Cgd* is the gate-drain capacitance per unit width, *W* is the channel width of M2, and *L* is the channel length of M2. The classical noise power spectral density (PSD) of 1/f noise of a MOS transistor given by,

$$S_v = \frac{K_f}{C_{ox}LWf} = \frac{N_f}{f} \qquad (7)$$

Where $k_f$ is a constant related to the interface state density of the MOSFET structure, *Cox* is the gate capacitance per unit area, *f* is the frequency, and $N_f = \frac{K_f}{C_{ox}LW}$ is denoted as flicker noise parameter. Considering CDS operation in Eqs.4 and Eqs.5 we get PDS of the 1/f noise of the source follower in a CMOS image sensor as,

$$S_{n,1/f} = P(1) = \beta_{CDS} N_f = \beta_{CDS} \frac{K_f}{C_{ox}LW} \qquad (8)$$

Now, we try to enhance the noise analysis step by step for modern submicron CMOS technology. In Eqs.6 it can be seen that the total input capacitance is the collection of floating diffusion capacitance and gate capacitance. To get an accurate model, any term depends on the geometry (*W, L*) of the input MOSFET must be included in the gate capacitance $C_G$. The extrinsic part of gate capacitance is known as the overlap capacitances. Overlap capacitances are proportional to the width *W* of the MOSFET and are not negligible in submicron CMOS technology.

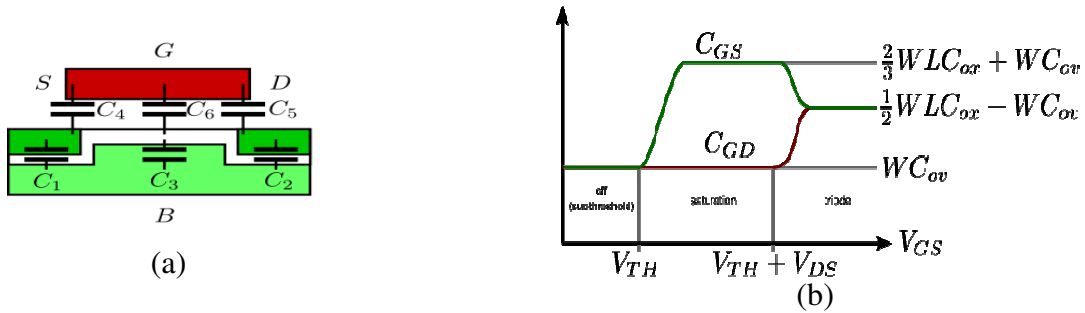

Figure 5: (a) Parasitic capacitances of MOSFET; (b) gate capacitances in different operation regions.

We consider gate-channel capacitance and overlap capacitances for noise analysis. In Figure 5(a) *C4* and *C5* are overlap capacitances and are only proportional to the width of the transistor. *C6* is the gate-channel capacitance and its total value is splited between drain and source in a way that depends on the region of operation of the MOSFET. Usually, we define the normalized capacitances *Cov* as Overlap capacitance per unit length (fF/μm), *Cox* as Gate to channel capacitance per unit area (fF/μm2), Gate-source capacitance ($C_{GS}$) and gate-drain capacitance ($C_{GD}$) have a base value of the overlap capacitance *CovW*. To that the gate to channel capacitance *CoxWL* according to the region of operation needs to be added. In-pixel source follower operates in saturation region. By considering improved model of gate capacitance for saturation region, the total capacitance at the input of the source follower is (see Figure 5(b)),

$$C_{total} = C_{FD} + C_G = C_{FD} + C_{GS} + C_{GD} = C_{FD} + 2C_{OV}W + \frac{2}{3}C_{ox}WL \quad (9)$$

We can write the mean-square 1/f noise charge as,

$$\overline{Q_{n,1/f}^2} = S_{n,1/f}C_{total}^2 = \beta_{CDS}\frac{K_f}{C_{ox}LW}(C_{FD} + 2C_{OV}W + \frac{2}{3}C_{ox}WL)^2 \quad (10)$$

As extrinsic capacitances are proportional to the width of the MOSFET, so we will try to find the optimum W for each L. Here geometry dependent part is,

$$f(W,L) = \frac{1}{LW}(C_{FD} + 2C_{OV}W + \frac{2}{3}C_{ox}WL)^2 \quad (11)$$

Minima of $f(W,L)$ can be obtained with optimal *W* and *L*. $f(W,L)$ will have extrema in the set of [*Wmin, Wmax*] X [*Lmin, Lmax*] where *Wmin* and *Lmin* are the minimum gate dimensions allowed by the technology. We can use simplified gradient descent as a first-order optimization technique. The local and global minima of $f(W,L)$ are correspond to the points that nullify the gradient,

$$\vec{\nabla}f(W,L) = \left(\frac{\partial f}{\partial W}, \frac{\partial f}{\partial L}\right) \quad (12)$$

Then, to find the point where $\frac{\partial}{\partial W}f(W,L)$ nullify we got the optimum width $W_{opt}$ for minimum noise charge for each *L* is,

$$\frac{\partial}{\partial W}f(W,L) = 0 \Leftrightarrow W_{opt} = \frac{C_{FD}}{2C_{OV} + \frac{2}{3}C_{ox}L} \quad (13)$$

For classical MOSFET model $\overline{Q_{n,1/f}^2} = S_{n,1/f}C_{total}^2 = \beta_{CDS}\frac{K_f}{C_{ox}LW}(C_{FD} + C_G)^2$ and $C_G = C_{ox}LW$.

From there, $\overline{Q_{n,1/f}^2} = \beta_{CDS}K_f\left[\frac{(C_{FD}+C_G)^2}{C_G}\right]$, whose minimum is reached for $C_G = C_{G,opt} = C_{FD}$. This is called capacitive matching. By applying capacitive matching and replacing *W* with $W_{opt}$ in Eqs.10 we can write optimum noise power $\overline{Q_{n,1/f,opt}^2}$ as:

$$\overline{Q_{n,1/f,opt}^2} \approx \frac{8}{3}\beta_{CDS}K_fC_{FD}\left(1 + \frac{3C_{OV}}{C_{ox}L}\right) \quad (14)$$

The plot in Figure 6 derived from Eqs.14 for n-channel MOSFET in Tower's 180nm technology (TSL018), demonstrates the impact of *L* on the optimum 1/f noise charge with optimum *W*.

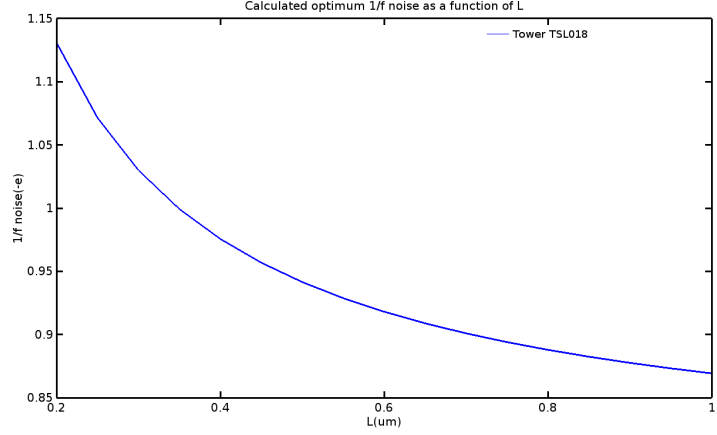

Figure 6: Calculated 1/f noise charge vs L for n-MOSFET source follower.

For this illustration we can say intuitively that as $L$ increases the 1/f noise decreases more rapidly than the increase in gate capacitance. Figure 7 shows the variation of the expression of Eqs.10 as a function of W and L of the source follower for Tower's TSL018 CMOS technology with $C_{OV} = 0.35\text{fF}/\mu\text{m}$, $C_{ox} = 5.02\text{fF}/\mu\text{m}^2$, $C_{FD} = 1.2\text{fF}$, and $K_f = 10^{-25}\text{V}^2\text{F/s}$ . As mentioned earlier to achieve precise FPN cancelation as an example with $\omega_c T_0 \cong 7$, the settling error is significantly low and from Figure 3 and Eqs.5 we consider $\beta_{CDS} \cong 5$.

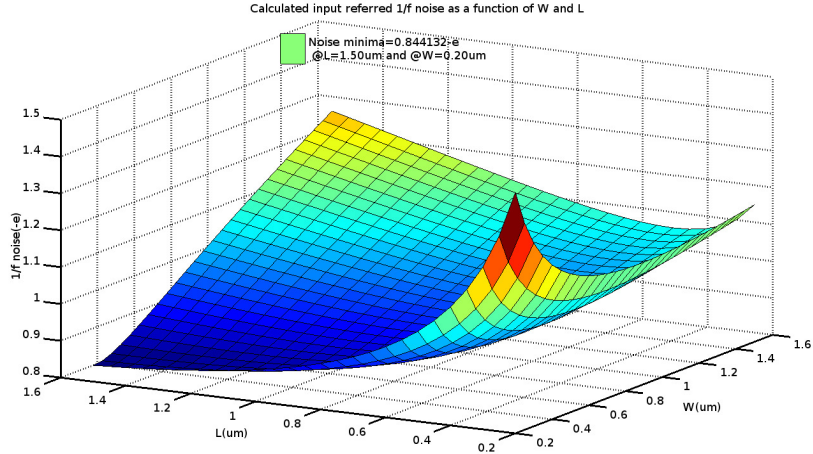

Figure 7: Calculated 1/f noise of SF as a function of W and L for Tower 180nm technology

Similarly we also derive equivalent 1/f noise charge equation of the source follower with classical capacitor model $C_{total} = C_{FD} + C_{GS} + C_{GD} = C_{FD} + C_{gs}WL + C_{gd}W$ to compare with the above. We also analytically calculate the noise power contribution for same available technology. Figure 8 shows the variation of that noise expression as a function of W and L of the source follower for a process technology with $C_{gs} = 3.4\text{fF}/\mu\text{m}^2$, $C_{gd} = 0.4\text{fF}/\mu\text{m}$, $C_{ox} = 5.02\text{fF}/\mu\text{m}^2$, $C_{FD} = 0.8\text{fF}$, $K_f = 10^{-25}\text{V}^2\text{F/s}$ and $\beta_{CDS} \cong 5$.

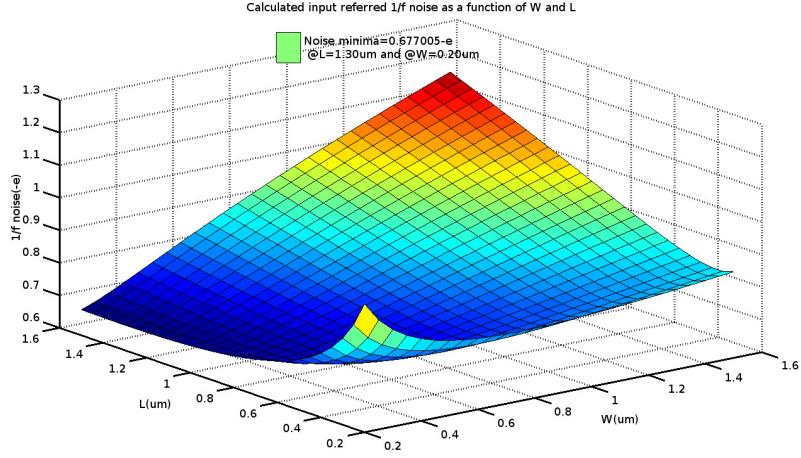

Figure 8: Calculated 1/f noise of SF as a function of W and L with above parameters

It is now safe to conclude that the best 1/f noise performance can be obtained by optimizing the layout to have smallest $C_{FD}$ and using lowest *W* with optimum *L*. Using enhanced capacitor model we observed that traditional smaller MOSFET approach makes the 1/f noise performance worst compare to the noise with classical capacitor model. In the above analysis of extended 1/f noise optimization for submicron technology included intrinsic and extrinsic gate capacitance but still did not take into account the bias dependent component. In deep submicron technologies this dependence can be relevant especially during transitions between different inversion states. It could be a good next step to extend the analysis further by including the bias dependent component with advanced noise model used in today's circuit simulators .

### 3.2  1/f Noise analysis including current source transistor

The above noise analysis did not include the current source transistor (M3 in Figure 4). In this point, we decided to analytically examine the contribution of the current source transistor in total 1/f noise before proceed with the enhanced noise model. Considering current source transistor for 1/f noise analysis total noise PSD is given by [6],

$$S_{n,1/f} = \frac{1}{f}\left[N_{f2} + \left(\frac{g_{m3}}{g_{m2}}\right)^2 N_{f3}\right] \tag{15}$$

Here $N_{f2}$ and $N_{f3}$ are the flicker noise parameter of M2 and M3. The transconductances $g_{m2}$ and $g_{m3}$ for M2 and M3 transistor is,

$$g_{m2} = \sqrt{2K\frac{W_2}{L_2}I_d} \text{ and } g_{m3} = \sqrt{2K\frac{W_3}{L_3}I_d} \tag{16}$$

Where $K = \mu_n C_{ox}$ and $I_d$ is the drain current through M2 and M3. $W_2$ and $W_3$ are the widths of M2 and M3. $L_2$ and $L_3$ are the lengths of M2 and M3 transistors. Then, the mean-square 1/f noise charge can be expressed as,

$$\overline{Q_{n,1/f}^2} = S_{n,1/f}C_{total}^2 = \frac{1}{f}\left[N_{f2} + \left(\frac{g_{m3}}{g_{m2}}\right)^2 N_{f3}\right](C_{FD} + C_{gs}WL + C_{gd}W)^2 \tag{17}$$

Replacing the expression of $g_{m2}$, $g_{m3}$, $N_{f2} = \frac{K_f}{C_{ox}W_2L_2}$ and $N_{f3} = \frac{K_f}{C_{ox}W_3L_3}$ we get,

$$\overline{Q_{n,1/f}^2} = \frac{K_f}{fC_{ox}W_2L_2}\left[1 + \left(\frac{L_2}{L_3}\right)^2\right](C_{FD} + C_{gs}WL + C_{gd}W)^2 \tag{18}$$

$$\overline{Q_{n,1/f}^2} = \beta_{CDS}\frac{K_f}{C_{ox}W_2L_2}\left[1 + \left(\frac{L_2}{L_3}\right)^2\right](C_{FD} + C_{gs}WL + C_{gd}W)^2 \tag{19}$$

From Eqs.19 we can see the geometry dependent part has three design variables ($W_2$, $L_2$, $L_3$) unlike previous analysis. That is,

$$f(W_2, L_2, L_3) = \frac{K_f}{C_{ox}W_2L_2}\left[1 + \left(\frac{L_2}{L_3}\right)^2\right](C_{FD} + C_{gs}WL + C_{gd}W)^2 \tag{20}$$

To find optimum value of $W_2, L_2 \text{ and } L_3$, analytically solving the Eqs.20 is complex in nature, so we numerically calculated the expression and plot in Figure 9 to see the effect of the current source transistor (M3) on 1/f noise. We selected $W_2$ as technology given minimum $W_{min}$=0.22µm for Tower 180nm technology. Figure 9 shows the length of the current source transistor (M3) in CMOS pixel has little effect on the overall 1/f noise and can be designed freely according to the current load requirement. May be a further detail analysis would be needed to have more clear view on the 1/f noise contribution of the current source transistor. We put that out-of-scope for this paper as in-pixel source follower is the main focus.

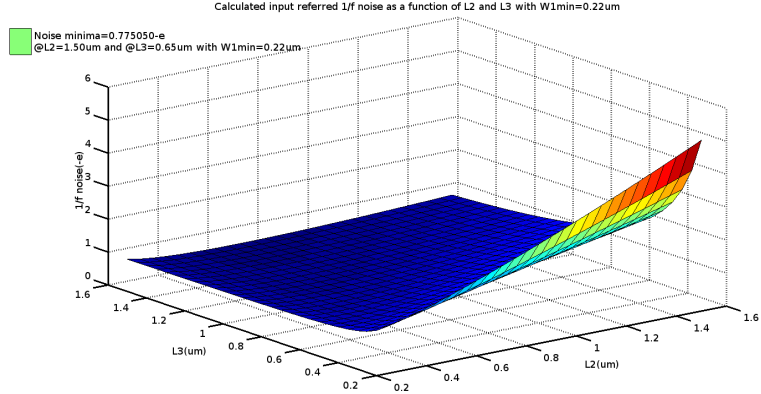

Figure 9: Calculated 1/f noise as a function of L1 and L2 with above parameters

### 3.3 1/f noise analysis for deep submicron technology with BSIM simulation model

Most of the current CMOS image sensor technology use BSIM3 or BSIM4 model [11] for circuit simulation. It is practical and accurate to investigate the noise model, used in BSIM3 or BSIM4. When the device is biased in saturation, the channel can be divided into the pinch-off and the non-pinch-off regions. The noise model can thus includes the noise equations for the two channel regions. The drain current noise density is defined in BSIM3v3 model as,

$$S_{I_d} = \frac{q^2 KT I_d \mu}{\gamma f^{EF} L^2 C_{ox} 10^8} \left[ NOIA.\ln\left(\frac{N_0 + 2.10^{14}}{N_l + 2.10^{14}}\right) + NOIB(N_0 - N_l) + \frac{NOIC}{2}(N_0^2 - N_l^2) \right] \\ + \frac{KT I_d^2 \Delta L_{clm}}{WL^2 f^{EF} 10^8} \cdot \frac{NOIA + NOIB.N_l + NOIC.N_l^2}{(N_l + 2.10^{14})^2} \quad (21)$$

$$S_{I_d} = \frac{q^2 KT I_d \mu}{\gamma f^{EF} L^2 C_{ox} 10^8} \cdot A + \frac{KT I_d^2 \Delta L_{clm}}{WL^2 f^{EF} 10^8} \cdot B \quad (22)$$

Here, 
$$A = NOIA.\ln\left(\frac{N_0 + 2.10^{14}}{N_l + 2.10^{14}}\right) + NOIB(N_0 - N_l) + \frac{NOIC}{2}(N_0^2 - N_l^2) \quad (23)$$

$$B = \frac{NOIA + NOIB.N_l + NOIC.N_l^2}{(N_l + 2.10^{14})^2} \quad (24)$$

Where, *NOIA*, *NOIB*, and *NOIC* are the noise parameters, which have empirical values according to the specific technology. *EF* is the flicker noise frequency exponent, $N_o$ is the charge density at the source side, $N_l$ is the charge density at the drain end, $\Delta L_{clm}$ is the channel length reduction due to channel length modulation.
The input referred noise voltage spectral density is define as,

$$S_{V_g} = \frac{S_{I_d}}{g_m^2} \quad (25)$$

When the effects of the carrier velocity saturation in the channel is negligible, the MOSFET's transconductance is presented as,

$$g_m^2 = 2\mu C_{ox}\left(\frac{W}{L}\right) I_d \quad (26)$$

Where, Cox is the oxide capacitance per unit area of the MOSFET. From Eqs.22 and Eqs.26 if we replace $S_{I_d}$ and $g_m^2$ in Eqs.25 we get,

$$S_{V_g} = \frac{q^2 KT.A}{\gamma f^{EF} C_{ox}^2 . 2.10^8} \cdot \frac{1}{WL} + \frac{KT \Delta L_{clm}.B}{\mu f^{EF} C_{ox}. 2.10^8} \cdot \frac{I_d}{W^2 L} \qquad (27)$$

From above expression the absolute minima of $S_{V_g}$ can be searched by setting the bias current at minimum.
Where,

$$I_d \geq I_{min} = R_{min} I_s \qquad (28)$$

Eqs.28 is the current for which noise equation is valid. $R_{min}$ is the minimum value of inversion coefficient and n is the sub-threshold slope.
$I_s$ is a reference current given by[12] as,

$$I_s = 2n\mu C_{ox} \frac{W}{L} \left(\frac{KT}{q}\right)^2 \qquad (29)$$

Replacing $I_d$ with $I_{min}$ in Eqs.27 we get,

$$S_{V_g} = \frac{q^2 KT.A}{\gamma f^{EF} C_{ox}^2 . 2.10^8} \cdot \frac{1}{WL} + \frac{(KT)^3 \Delta L_{clm}.B}{q^2 f^{EF}. 2.10^8} \cdot \frac{2n R_{min}}{WL^2} \qquad (30)$$

$$S_{V_g} = X \cdot \frac{1}{WL} + Y \cdot \frac{1}{WL^2} \qquad (31)$$

In Eqs.31, X and Y are considered as constant with respect to the circuit design parameter. All the associated value in X and Y are either absolute constant or technology dependent constant parameters.
From Eqs.17 we can rewrite the mean-square 1/f noise charge as,

$$\overline{Q_{n,1/f}^2} = S_{V_g} C_{total}^2 = \left(X \cdot \frac{1}{WL} + Y \cdot \frac{1}{WL^2}\right) \cdot \left(C_{FD} + 2C_{OV}W + \frac{2}{3}C_{ox}WL\right)^2 \qquad (32)$$

Then, in this above noise expression, the geometry dependent part can be written as,

$$f(W,L) = f_1(W,L) + f_2(W,L) \qquad (33)$$

As for the simplicity and to find out the local minima with initial guess we consider the weight of the design dependent function $f_n(W,L)$ is negligible. Here *n* is 1 or 2 in above expression.
Where,

$$f_1(W,L) = \frac{1}{WL}\left(C_{FD} + 2C_{OV}W + \frac{2}{3}C_{ox}WL\right)^2 \qquad (34)$$

And,

$$f_2(W,L) = \frac{1}{WL^2}\left(C_{FD} + 2C_{OV}W + \frac{2}{3}C_{ox}WL\right)^2 \qquad (35)$$

To find the point where $\frac{\partial}{\partial L} f(W,L)$ nullify, we solved this for $W \neq \infty$ and $L > 0$ and we get the optimum length $L_{opt}$ for minimum noise charge for each W is,

$$L = \frac{C_{FD}}{\frac{2}{3}C_{ox}W} + \frac{2C_{OV}}{\frac{2}{3}C_{ox}} \qquad (36)$$

Above analysis in this section provides a technique to find the optimum design point of the source follower to get minimum 1/f noise in CMOS image sensor. It also analyzes the effect of the current source transistor on the overall 1/f noise. This concludes that the current source transistor (M3) in CMOS pixel has little effect on the overall 1/f noise and can be designed freely according to the current load requirement.

## 4. ANALYSIS OF TOTAL NOISE OF THE SOURCE FOLLOWER (SF) FOR DEEP SUBMICRON TECHNOLOGY WITH BSIM SIMULATION MODEL TO OPTIMIZE

To study the total noise contribution of the source follower of a typical CMOS pixel we need to consider white noise or thermal noise along with the 1/f noise. In general white noise is a sum of series and parallel white noise contribution. For source follower, series noise is mainly thermal noise and parallel noise is shot noise or leakage noise. We can safely ignore shot noise for noise optimization by circuit design as it not depends on the circuit design parameters. As mentioned earlier that in current-time most of the current CMOS image sensor technology use BSIM3 or BSIM4 model for circuit simulation so it is practical to investigate the noise model, used in BSIM3 or BSIM4.

Here we used BSIM3v3 models for analysis. There are total four combinations of noise models depending on the particular 1/f noise and thermal noise model. The model parameter NOIMOD is used to select the particular combination (See Table 1).

Table 1: Combination of noise models in BSIM3v3.

| Noise parameter | Noise models | |
|---|---|---|
| NOIMOD | Flicker or 1/f noise model | Thermal noise model |
| 1 | Spice2 | Spice2 |
| 2 | BSIM3 | BSIM3 |
| 3 | BSIM3 | Spice2 |
| 4 | Spice2 | BSIM3 |

We found that most of the current image sensor technology use NOIMOD=3 combination so, we did the analysis for these noise models, indicated in Table 1(marked in yellow). The complete analysis to optimize the BSIM3 1/f noise model has been discussed in earlier section. Here we discuss the thermal noise (white series noise) and finally the total noise optimization. In Spice2 simulation model thermal noise current density is define as:

$$\frac{S_I}{\Delta f} = \frac{8}{3} KT |g_m + g_d + g_{mb}| \tag{37}$$

If source and substrate are in same potential then $Vsb$=0 so, $g_{mb} = 0$. In saturation region $g_d$ is negligible and $g_m \gg g_d$. So, thermal noise current density is,

$$\frac{S_I}{\Delta f} = \frac{8}{3} KT |g_m| \tag{38}$$

Source follower's voltage noise spectral density in strong inversion and in saturation is given by:

$$S_V = \frac{S_I}{g_m^2} = \frac{\frac{8}{3} KT |g_m|}{g_m^2} = \frac{8}{3} \frac{KT}{g_m} \tag{39}$$

Here $S_I$ is the thermal current noise spectral density and $g_m$ is the transconductance of the source follower. We can write the mean-square white noise charge as:

$$\overline{Q_w^2} = \frac{8}{3} \frac{KT}{g_m} (C_{FD} + C_G)^2 \frac{1}{\tau_p} \tag{40}$$

Where total capacitor at gate of the SF is $C_{total} = C_{FD} + C_G$ and $\tau_p$ is the time parameter, referred to peaking time. After replacing the expression of $g_m$ we rewrite,

$$\overline{Q_w^2} = \frac{8KTL}{3.\sqrt{2\mu_0}} \cdot \frac{(C_{FD} + C_G)^2}{\sqrt{C_{ox}WL}} \cdot \frac{1}{\sqrt{I_d}} \cdot \frac{1}{\tau_p} \tag{41}$$

For simplicity we assumed $C_G = C_{ox}WL$,
Then,

$$\overline{Q_w^2} = \frac{8KTL}{3.\sqrt{2\mu_0}} \cdot \frac{(C_{FD} + C_G)^2}{\sqrt{C_G}} \cdot \frac{1}{\sqrt{I_d}} \cdot \frac{1}{\tau_p} = K_w \cdot \frac{(C_{FD} + C_G)^2}{\sqrt{C_G}} \cdot \frac{1}{\sqrt{I_d}} \cdot \frac{1}{\tau_p} \tag{42}$$

From Eqs.30 we got the 1/f noise voltage spectral density in BSIM3 model as:

$$S_{V_g} = \frac{q^2 KT.A}{\gamma f^{EF} C_{ox}^2.2.10^8} \cdot \frac{1}{WL} + \frac{KT \Delta L_{clm}.B}{\mu f^{EF} C_{ox}.2.10^8} \cdot \frac{I_d}{W^2 L} \tag{43}$$

After simplifying with replacing $C_{ox}WL = C_G$ and $S_{V_g}$, the mean-square 1/f noise charge density of the SF is expressed as,

$$\overline{Q_{n,1/f}^2} = S_{V_g} \cdot (C_{FD} + C_G)^2 = \left[ K_{f1} \cdot \frac{(C_{FD} + C_G)^2}{C_G} + K_{f2} \cdot \frac{I_d(C_{FD} + C_G)^2}{C_G} \right] \quad (44)$$

Where, $K_{f1} = \frac{q^2 KT.A}{\gamma f^{EF} C_{ox}.2.10^8}$ and $K_{f2} = \frac{KT \Delta L_{clm}.B}{\mu f^{EF} W.2.10^8}$.

The total noise contribution of the source follower is the sum of the white noise and 1/f noise. From Eqs.42 and Eqs.44,

$$\overline{Q_{total}^2} = K_w \cdot \frac{(C_{FD} + C_G)^2}{\sqrt{C_G}} \cdot \frac{1}{\sqrt{I_d}} \cdot \frac{1}{\tau_p} + K_{f1} \cdot \frac{(C_{FD} + C_G)^2}{C_G} + K_{f2} \cdot \frac{I_d(C_{FD} + C_G)^2}{C_G} \quad (45)$$

In case of the constant pecking time the total noise is a function of $I_d$ and $C_G$. The optimum $I_d$ and $C_G$ to achieve minima of $\overline{Q_{total}^2}$ can be found using gradient method by nullify its partial derivatives with respect to $I_d$ and $C_G$.

For given $C_G$,

$$\frac{\partial \overline{Q_{total}^2}}{\partial I_d} = \frac{\partial}{\partial I_d} \left[ K_w \cdot \frac{(C_{FD} + C_G)^2}{\sqrt{C_G}} \cdot \frac{1}{\sqrt{I_d}} \cdot \frac{1}{\tau_p} + K_{f1} \cdot \frac{(C_{FD} + C_G)^2}{C_G} + K_{f2} \cdot \frac{I_d(C_{FD} + C_G)^2}{C_G} \right] \quad (46)$$

$$= \frac{\partial}{\partial I_d} \left[ M \cdot \frac{1}{\sqrt{I_d}} + K_{f1} \cdot \frac{(C_{FD} + C_G)^2}{C_G} + N \cdot I_d \right]$$

Where, $M = K_w \cdot \frac{(C_{FD}+C_G)^2}{\sqrt{C_G}} \cdot \frac{1}{\tau_p}$ and $N = K_{f2} \cdot \frac{(C_{FD}+C_G)^2}{C_G}$.

After solving Eqs.46 and nullify it, we get the optimum bias current as,

$$I_d = I_{d,opt} = \left( \frac{M}{2N} \right)^{\frac{2}{3}} \quad (47)$$

Then,

$$I_{d,opt} = \left( \frac{K_w \cdot \frac{(C_{FD}+C_G)^2}{\sqrt{C_G}} \cdot \frac{1}{\tau_p}}{2 K_{f2} \cdot \frac{(C_{FD}+C_G)^2}{C_G}} \right)^{\frac{2}{3}} = \left( \frac{K_w}{K_{f2} \cdot \tau_p} \cdot \sqrt{C_G} \right)^{\frac{2}{3}} \quad (48)$$

On the other hand for a given bias current total noise can be minimized at an optimum gate capacitance $C_G = C_{G,opt}$. That can be calculated by gradient minimization method as is used earlier. We take the partial differentiation of total noise charge density with respect to $C_G$ and nullify then we get a cubic equation below.

$$3A\left(\sqrt{C_G}\right)^3 + 2B\left(\sqrt{C_G}\right)^2 - AC_{FD}\sqrt{C_G} - 2BC_{FD} = 0 \quad (49)$$

Where, $A = K_w \cdot \frac{1}{\sqrt{I_d}} \cdot \frac{1}{\tau_p} = $ constant and $B = K_{f1} + K_{f2}I_d = $ constant. With the condition, defined in Eqs.28 and Eqs.29 as,

$$I_d \geq I_{min} = R_{min} I_s \quad (50)$$

$$I_s = 2n\mu C_{ox} \frac{W}{L} \left(\frac{KT}{q}\right)^2 = 2n\mu C_{ox} WL \left(\frac{KT}{qL}\right)^2 = 2n\mu C_G \left(\frac{KT}{qL}\right)^2 \quad (51)$$

If we replace the value of $I_s$ from Eqs.51 in Eqs.50, we get the condition for gate capacitance as,

$$C_G \leq \frac{I_d}{2n\mu R_{min} \left(\frac{KT}{qL}\right)^2} \quad (52)$$

For Eqs.49 we can get two limit solutions. For no white noise component, $K_w = 0$, so $A = 0$ then the solution is $C_G = C_{G,opt} = C_{FD}$. On the other hand for no 1/f noise component $K_{f1} = 0$ and $K_{f2} = 0$, so $B = 0$ in Eqs.49 then the solution is $C_G = C_{G,opt} = \frac{C_{FD}}{3}$.

# 5. DESIGN PARAMETERS AND SIMULATION RESULTS

The total noise model also depends on the in-pixel source follower's bias current (From Eqs.45) in contrast to the earlier noise model for large device. Therefore, it is understood that we can only find design conditions for relative minima. When optimum bias current is higher than the minimum allowable current described in Eqs.50, 1/f noise become dominating compare to the white series noise component. If the optimum current $I_{d,opt}$ is lower than the minimum current $I_{min}$ it must be set as $I_d = I_{min}$ according to the condition in Eqs.50. If we express the ratio of $I_{min}$ and $I_{d,opt}$ as $\rho = \frac{I_{min}}{I_{d,opt}}$ then we can represent the quasi-optimum gate capacitance as,

$$C_{G,Qopt} = C_{FD}\left(\frac{1+3\rho}{3+\rho}\right) \tag{53}$$

When $\rho = 1$ means $I_{d,opt} = I_{min}$ at which we can see from section 4 that the total noise has no white noise component and 1/f noise is dominant. So, $\overline{Q_w^2} = \overline{Q_{n,1/f}^2}$ and the optimum gate capacitance will be $C_G = C_{G,opt} = C_{FD}$. It can be observed that if drain current is increased further, the quasi-optimum capacitor ratio tends toward $\frac{C_G}{C_{FD}} = \frac{1}{3}$ as the white or thermal noise becomes dominant. So considering bias dependency it is understood that for a given current $I_d \geq I_{min} = R_{min}I_s$ relative minima of the total noise will be obtained with optimum $C_G$. From above study, bias current can be expressed as,

$$I_d = \frac{C_G}{C_{FD}} \cdot I_{min} \tag{54}$$

Rearranging the Eqs.53 and replacing the expression $\rho = \frac{I_{min}}{I_{d,opt}}$ we write,

$$\frac{I_{min}}{I_{d,opt}} = \left(\frac{1-3\alpha}{\alpha-3}\right) \tag{55}$$

Where, $\alpha = \frac{C_G}{C_{FD}}$

From Eqs.54 and Eqs.55, after replacing $I_{min}$, we get,

$$\frac{I_d}{I_{d,opt}} = \alpha\left(\frac{1-3\alpha}{\alpha-3}\right) = \frac{C_G}{C_{FD}} \cdot \left(\frac{1-3 \cdot \frac{C_G}{C_{FD}}}{\frac{C_G}{C_{FD}} - 3}\right) \tag{56}$$

Eqs.56 can be used to find best $C_G$ as design parameter to minimize the total noise for specific bias condition. Figure 10 is the plot of Eqs.56. In the figure it can be noticed that at $I_d = I_{d,opt}$, the optimum gate capacitance is equal to the floating diffusion capacitance, $C_G = C_{G,opt} = C_{FD}$. When bias current is larger than its optimum value as $I_d \gg I_{d,opt}$, the best $C_G$ tends to $C_G \cong 3C_{FD}$ because $\overline{Q_{n,1/f}^2} \gg \overline{Q_w^2}$ whereas when $I_d \ll I_{d,opt}$, the $C_G$ tends towards $C_G \cong C_{FD}/3$ as expected because $\overline{Q_w^2} \gg \overline{Q_{n,1/f}^2}$.

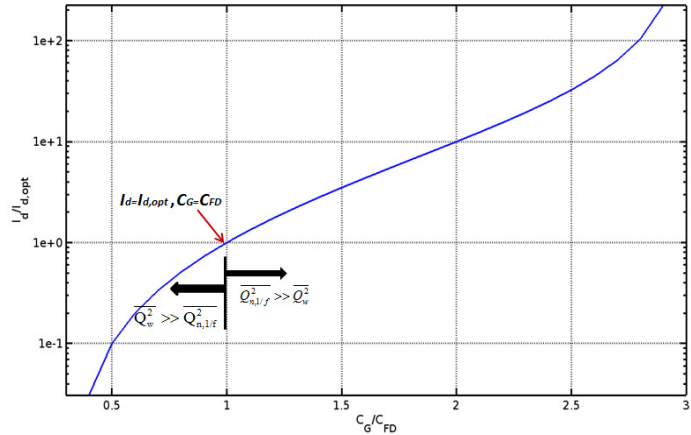

Figure 10: Plot of Eqs.70 from which optimum gate capacitance (CG) can be obtain to get minimum total noise

After analytical finding, simulation validations are performed with commercial technology to observe the bias dependency. Tower's 180nm technology (TSL018) has been used where MOSFET models are BSIM3v3.24.

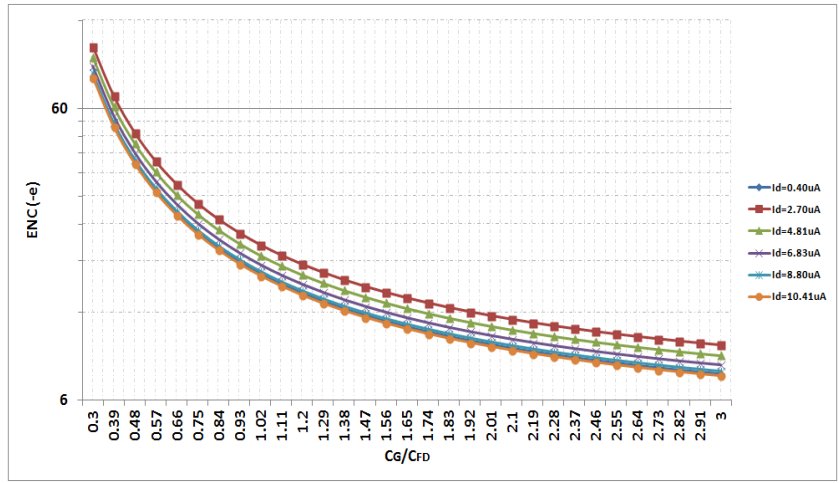

Figure 11: Simulated equivalent noise charge (ENC) vs capacitor matching ratio at different drain current of the source follower

The parametric simulation of CMOS pixel circuit is done to find out how total noise, generated from SF depends on the capacitor matching optimization with different biasing current (drain current). Figure 11 shows that the equivalent noise charge density not only depends on $C_G/C_{FD}$ ratio but also on the drain current of the source follower. Therefore, the biasing-current needs to be considered as an important design parameter to find the correct capacitor matching ratio and to get the optimized noise.

In Figure 12 we plotted capacitor matching ratio CG/CFD with respect to the bias current. It is noticed that to get a particular noise level we need to design different capacitor matching ration for different biasing current. In Figure 12 the particular noise level is used as 12e- equivalent noise charge (ENC).

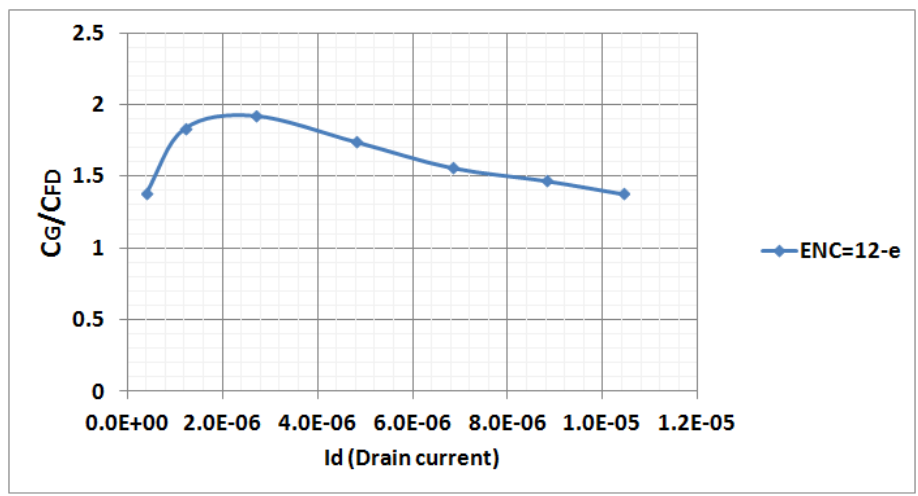

Figure 12: Simulated capacitor matching ratio at different drain current to get 12-e equivalent noise charge at the input of the source follower

## 6. CONCLUSION

In this work, CMOS in-pixel source follower's white noise and flicker (1/f) noise are analytically studied with BSIM3v3 models. Enhanced capacitor model for the gate capacitance of the source follower is considered in calculation of the equivalent noise power. Our detailed analysis shows the dependence of the 1/f noise on the geometrical size of the source follower is not limited to minimum channel length, compared to the classical approach to achieve the minimum 1/f noise. We derive the optimal gate dimensions (the width and the length) of the source follower that minimize the 1/f noise, and validate our results using numerical simulations. In the presence of both the white noise and 1/f noise, total noise is shown as bias dependent. It is also shown that optimum $C_G$ depends on the chosen bias condition but ranges as $\frac{C_{FD}}{3} \leq C_{G,opt} \leq C_{FD}$ (see Eqs.49). As a possible design steps, Eqs.56 is derived which can be used to find best gate capacitance $C_G$ as a design parameter to minimize the total noise for specific bias condition. Finally the parametric simulation of CMOS pixel circuit is done with Tower's 180nm technology (TSL018). This simulation demonstrated, how total noise, generated from SF, depends on the capacitor matching optimization with different biasing current (drain current) which also validated our analytical findings. In particular, it has been concluded that analytical study of the exact noise model, used in simulator, gives more detail and accurate predictive design parameter to achieve optimized noise performance of the CMOS in-pixel source follower.

## ACKNOWLEDGMENTS

This work is funded by IWT-Baekeland.